\documentclass[sigconf]{acmart}
\usepackage{graphicx}
\usepackage{latexsym}
\usepackage{amsmath}
\usepackage{booktabs}
\usepackage{multirow}
\usepackage[ruled,vlined]{algorithm2e}
\usepackage{svg}
\usepackage{amsthm,amsmath}
\usepackage{mathrsfs}
\usepackage{tabularx}
\usepackage{enumitem}

\AtBeginDocument{%
  }
\setcopyright{acmlicensed}

\copyrightyear{2024}
\acmYear{2024}
\setcopyright{acmlicensed}\acmConference[SIGIR '24]{Proceedings of the 47th International ACM SIGIR Conference on Research and Development in Information Retrieval}{July 14--18, 2024}{Washington, DC, USA}
\acmBooktitle{Proceedings of the 47th International ACM SIGIR Conference on Research and Development in Information Retrieval (SIGIR '24), July 14--18, 2024, Washington, DC, USA}
\acmDOI{10.1145/3626772.3657973}
\acmISBN{979-8-4007-0431-4/24/07}

\begin{document}
\newcolumntype{L}[1]{>{\raggedright\arraybackslash}p{#1}}
\newcolumntype{C}[1]{>{\centering\arraybackslash}p{#1}}
\newcolumntype{R}[1]{>{\raggedleft\arraybackslash}p{#1}}

\title{Behavior Pattern Mining-based Multi-Behavior Recommendation }


\author{Haojie Li}
\orcid{0009-0001-0863-9576}
\email{lihaojie@qust.edu.cn}
\affiliation{%
  \institution{School of Data Science, Qingdao University of Science and Technology}
  \city{Qingdao}
  \country{China}
}

\author{Zhiyong Cheng}
\orcid{0000-0003-1109-5028}
\email{jason.zy.cheng@gmail.com}
\affiliation{%
  \institution{School of Computer Science and Information Engineering, Hefei University of Technology}
  \city{Hefei}
  \country{China}
}

\author{Xu Yu}
\orcid{0000-0003-4913-5734}
\email{yuxu0532@upc.edu.cn}
\affiliation{%
  \institution{College of Computer Science and Technology, China University of Petroleum}
  \city{Qingdao}
  \country{China}
}

\author{Jinhuan Liu}
\orcid{0000-0002-1151-6040}
\email{liujinhuan.sdu@gmail.com}
\affiliation{%
  \institution{School of Data Science, Qingdao University of Science and Technology}
  \city{Qingdao}
  \country{China}
}

\author{Guanfeng Liu}
\orcid{0000-0001-8980-4950}
\email{guanfeng.liu@mq.edu.au}
\affiliation{%
  \institution{School of Computing, Macquarie University}
  \city{Sydney}
  \country{Australia}
}

\author{Junwei Du}
\authornote{Junwei Du is the corresponding author.}
\orcid{0000-0002-2909-2565}
\email{djwqd@163.com}
\affiliation{%
  \institution{School of Data Science, Qingdao University of Science and Technology}
  \city{Qingdao}
  \country{China}
}

\renewcommand{\shortauthors}{Haojie Li et al.}

\begin{abstract}

Multi-behavior recommendation systems enhance effectiveness by leveraging auxiliary behaviors (such as page views and favorites) to address the limitations of traditional models that depend solely on sparse target behaviors like purchases. 
Existing approaches to multi-behavior recommendations typically follow one of two strategies: some derive initial node representations from individual behavior subgraphs before integrating them for a comprehensive profile, while others interpret multi-behavior data as a heterogeneous graph, applying graph neural networks to achieve a unified node representation. 
However, these methods do not adequately explore the intricate patterns of behavior among users and items.
To bridge this gap, we introduce a novel algorithm called Behavior Pattern mining-based Multi-behavior Recommendation (BPMR). Our method extensively investigates the diverse interaction patterns between users and items, utilizing these patterns as features for making recommendations. We employ a Bayesian approach to streamline the recommendation process, effectively circumventing the challenges posed by graph neural network algorithms, such as the inability to accurately capture user preferences due to over-smoothing. Our experimental evaluation on three real-world datasets demonstrates that BPMR significantly outperforms existing state-of-the-art algorithms, showing an average improvement of $268.29\%$ in Recall@10 and $248.02\%$ in NDCG@10 metrics. 
The code of our BPMR is openly accessible for use and further research at \href{https://github.com/rookitkitlee/BPMR}{https://github.com/rookitkitlee/BPMR}.

\end{abstract}





\maketitle

\begin{figure}[t]
\setlength{\abovecaptionskip}{5pt}
\setlength{\belowcaptionskip}{-5pt}
\centering
\includegraphics[width=0.3\textwidth]{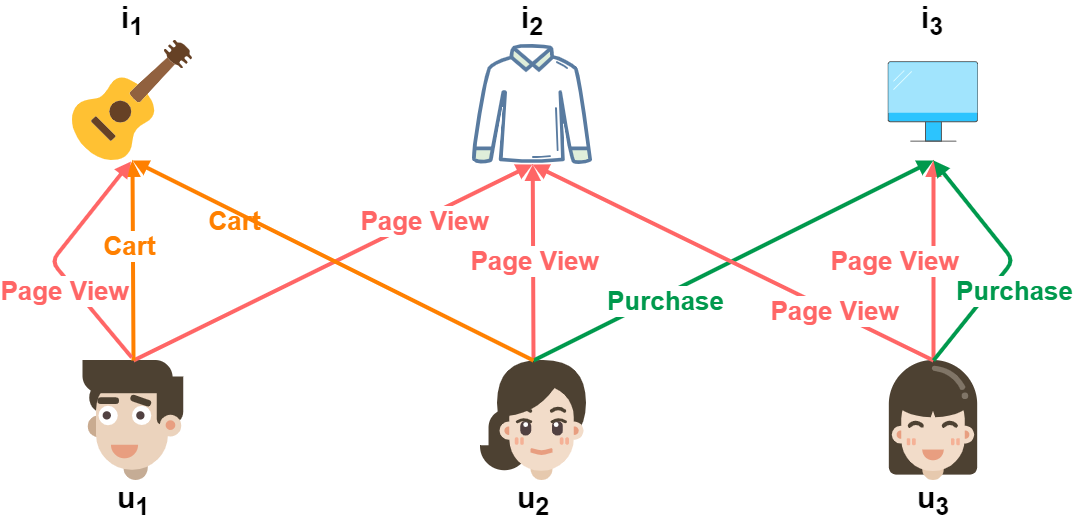}
\caption{The user-item interaction graph.}
\label{fig:label}
\end{figure}

\section{Introduction}
Recommendation systems effectively resolve information overload by accurately predicting individualized user preferences \cite{wuSurveyAccuracyorientedNeural2022}.
Model-based collaborative filtering methods \cite{yuAreGraphAugmentations2022,maoUltraGCNUltraSimplification2023,caiLightGCLSimpleEffective2023}, learning from user-item interactions, effectively construct representations of them, becoming central to recommendation technology.
Traditional recommendation approaches primarily focus on a single type of user behavior, such as purchases. 
However, in real-world applications, these singular behavior models encounter significant challenges due to data sparsity \cite{xuMultibehaviorSelfsupervisedLearning2023,zhangDenoisingPromptTuningMultiBehavior2023,zhangAlleviatingBehaviorData2023}.
Recent research increasingly emphasizes the importance of multi-behavior recommendation systems \cite{weiContrastiveMetaLearning2022,xuMultibehaviorSelfsupervisedLearning2023,yanCascadingResidualGraph2023}. 
These systems consider not only the target behavior but also a range of auxiliary behaviors, such as page views and favorites. 
Compared to focusing solely on the target behavior, auxiliary behaviors provide richer user interaction data, aiding in more accurately capturing user preferences and effectively addressing data sparsity issues in recommendation systems.

Current multi-behavior recommendation algorithms can be divided into two categories based on how they enhance the representation of target behaviors using auxiliary behaviors.
The first category is the decomposition-enhanced pattern, where multi-behavior interactions are divided into individual behavior subgraphs for separate embedding. 
This approach facilitates separate embedding learning for each behavior, subsequently enriching the target behavior's representation by integrating insights from auxiliary behaviors. 
For example, some methods use self-supervised learning to embed structural information of auxiliary behaviors into the target behavior representation \cite{weiContrastiveMetaLearning2022,guSelfsupervisedGraphNeural2022,xuMultibehaviorSelfsupervisedLearning2023}, 
while others explore the sequential or chain-like nature of behavioral relationships, aggregating this information to unearth more profound insights into user preferences
\cite{yanCascadingResidualGraph2023,chengMultiBehaviorRecommendationCascading2023}. 
The second category is the unified encoding pattern, which aims to unify the processing and integration of diverse user behaviors to generate node representations.
Methods in this category include the application of matrix collaborative filtering to handle interactions under multiple behaviors \cite{singhRelationalLearningCollective,chenEfficientHeterogeneousCollaborative2020},
and the extension of convolutional neural networks to incorporate diverse behavior information during message passing \cite{jinMultibehaviorRecommendationGraph2020,chenGraphHeterogeneousMultiRelational2021}.


Although current multi-behavior recommendation algorithms demonstrate effectiveness, they fall short in thoroughly investigating the intricate patterns of behavior that occur between nodes. 
Consider, for example, a complex interaction chain depicted in Figure 1, where a user $u_1$ views a page $i_2$, 
followed by another user $u_2$ viewing the same page and subsequently purchasing item $i_3$. 
This sequence "PageView followed by PageView and then a Purchase" between 
$u_1$ and $i_3$ exemplifies a nuanced behavior pattern that current algorithms struggle to decipher fully.

Algorithms based on the decomposition-enhanced pattern only represent single subgraphs and do not explore the interactions of different behaviors between nodes. 
Those based on the unified encoding pattern, while capable of processing multiple behaviors via graph neural networks, mix the interaction information of the current node with different neighbors and behaviors, losing the detailed features of a specific behavior pattern. 
However, in some cases, predicting graph structures typically requires focusing only on certain specific structural features \cite{weiBoostingGraphContrastive, yuFindingDiversePredictable2022}.
Furthermore, most existing current multi-behavior recommendation algorithms rely on graph collaborative filtering, facing an over-smoothing issue with graph neural networks that hinders the accurate capture of unique user preferences
\cite{liuDeeperGraphNeural2020,zhouDeeperGraphNeural2020,xiaGraphlessCollaborativeFiltering2023}.


To overcome the limitations identified in existing multi-behavior recommendation systems, we introduce the Behavior Pattern mining-based Multi-behavior Recommendation (BPMR) algorithm. 
Our approach delves into the complex interplays of behavior patterns that exist between users and items, leveraging these insights to inform our recommendations. 
By calculating the likelihood of various behavior patterns occurring alongside target behaviors, we utilize the Bayesian method to refine our item recommendations.
The contributions of our work are summarized as follows:

\begin{itemize}[leftmargin=10pt]
\item
The BPMR algorithm conducts a thorough analysis of the diverse behavior patterns among nodes, applying these insights directly to enhance target behavior recommendations. By acknowledging the varied connections and the multi-hop relationships between nodes, our algorithm surpasses the limitations of traditional graph neural networks, which often struggle to represent intricate behavior patterns accurately.

\item
Compared to existing frameworks, our algorithm offers a more streamlined approach to model training, significantly enhancing efficiency. This simplicity also circumvents the common issue in graph collaborative filtering algorithms where distinct user preferences are not accurately captured, thereby increasing the precision of our recommendations.

\item 
We conducted experiments on three real-world datasets, using two evaluation metrics, and validated its robustness in scenarios with sparse target behavior and noisy auxiliary behavior.
\end{itemize}

\section{PRELIMIARIES}

\noindent\textbf{Multi-behavior Recommendation.}
In multi-behavior recommendation systems, the user and item sets are denoted as $\mathcal{U}(u \in \mathcal{U})$ and $\mathcal{I}(i \in \mathcal{I})$, respectively. Here, $|\mathcal{U}|$ and $|\mathcal{I}|$ indicate the total number of users and items. 
Additionally, the set of behaviors occurring between users and items is represented as $\mathcal{B}(b \in \mathcal{B})$, where $|\mathcal{B}|$  denotes the number of distinct behaviors, with the constraint that $|\mathcal{B}| \ge 2$. 
We define the ${|\mathcal{B}|}$-th type of behavior $b_{|\mathcal{B}|}$ as the target behavior and the other behaviors as auxiliary behaviors. 
The interaction matrix corresponding to the $b$-th type of behavior between users and items is defined as $\textbf{E}^b$, where 
\begin{equation}
\footnotesize
\textbf{E}_{u,i}^b = 
\begin{cases}
1, \text {If $u$ has interacted with $i$ under behavior $b$;} \\
0, \text {otherwise.}
\end{cases}
\end{equation}

The task of multi-behavior recommendation is formulated as:

\begin{itemize}[leftmargin=10pt]
\item \textbf{Input:} 
The interaction data encompassing all types of behaviors $\{\textbf{E}^1, \textbf{E}^2, \cdots, \textbf{E}^{|\mathcal{B}|}\}$, are collected from sets $\mathcal{U}$ and $\mathcal{I}$.
\item \textbf{Output:} 
A recommendation model that estimates the probability of users interacting with items in the target behavior, given their interactions across multiple types of behaviors.
\end{itemize}

\noindent\textbf{Behavior Pattern.}
In multi-behavior recommendation systems, a multi-behavior interaction path $p$ of length $l$ is defined as 
$n_1 \stackrel{b_1} \rightarrow n_2  \stackrel{b_2} \rightarrow ... \stackrel{b_l} \rightarrow n_{l+1}$, where $n_i \in \mathcal{U} \cup \mathcal{I}$ and $b_i \in \mathcal{B}$. 
It describes a composite relation $R=b_1 \circ b_2 \circ ... \circ b_l$ between nodes 
$n_1$ and $n_{l+1}$, where $\circ$ represents the combinatorial operator on the relation.
We define $R$ as a behavior pattern from node $n_1$ to node $n_{l+1}$, and $p$ as a path of the $R$-type behavior pattern from node $n_1$ to node $n_{l+1}$.

\section{METHODOLOGY}

\subsection{Behavior pattern mining}

In this section, our goal is to uncover behavior patterns between users and items. 
Taking an e-commerce platform as an example, potential interactions between users and items under various behaviors are illustrated in Figure 2. 
 This figure showcases potential interactions under various behaviors and their implications for understanding user intent.
\begin{itemize}[leftmargin=10pt]
\item \textbf{Shared Interests through Page Views (Figure 2(a))}: When users $u_1$ and $u_2$ both view item $i_1$, it signifies a common interest. The subsequent page view of item $i_2$ by $u_2$
  may imply a potential purchase interest for $u_1$
  as well, suggesting a transitive interest pattern.

\item \textbf{Direct Preliminary Purchase Intent (Figure 2(b))}: The direct interaction of $u_3$ with $i_3$ through a page view indicates a straightforward preliminary purchase intent, which is deemed more reliable than the inferred intents seen in Figure 2(a).

\item \textbf{Enhanced Interest Similarity through Cart Actions (Figure 2(c))}: The action of both $u_4$ and $u_5$
  adding item $i_4$ to their carts signals a stronger interest similarity than mere page views, as in Figure 2(a). Consequently, when $u_5$ views another item $i_5$
 , it strongly suggests a possible purchase intent for $i_4$
  as well.

\item \textbf{Strengthened Purchase Intent through Multiple Cart Actions (Figure 2(d))}: The similarity and likely purchase intent are further amplified when $u_6$ and $u_7$ have multiple items in common in their carts. If $u_7$ then views item $i_8$, it significantly boosts the likelihood of $u_6$'s interest in $i_8$.
\end{itemize}

From this exploration, we distill four crucial insights regarding user-item behavior patterns:
\begin{itemize}[leftmargin=10pt]
\item \textbf{Predictive Basis}: These patterns can reliably forecast future interactions with target items, serving as a predictive foundation.
\item \textbf{Pattern Length Relevance}: The effectiveness of these predictions varies with the length of the observed behavior patterns.
\item \textbf{Influence of Behavior Types}: Different types of interactions (such as page views vs. cart actions) have distinct impacts on predictive accuracy.
\item \textbf{Significance of Pattern Paths}: The number of behavior pattern paths plays a crucial role in formulating accurate target behavior recommendations.
\end{itemize}

\begin{figure}[t]
\setlength{\abovecaptionskip}{5pt}
\setlength{\belowcaptionskip}{-5pt}
\centering
\includegraphics[width=0.4\textwidth]{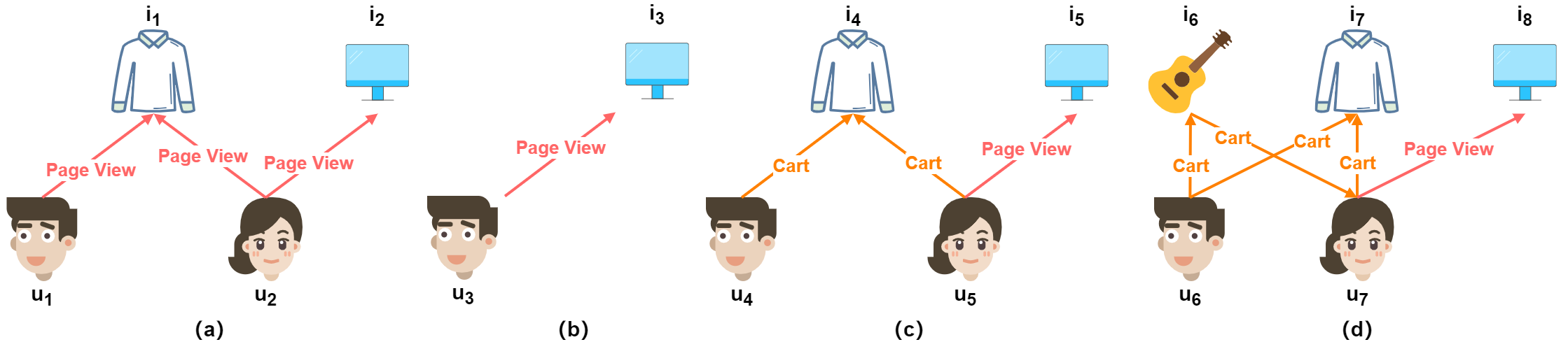}
\caption{Different types of behavior patterns between nodes.}
\label{fig:label}
\end{figure}

Drawing inspiration from the previously discussed scenarios, we devise a method to discern behavior patterns between users and items, focusing on two key dimensions: the pattern's length and the specific combination of behaviors it encompasses.
We use the number of paths corresponding to these behavior patterns as features for targeted behavior recommendations.
We define $\mathcal{S}_l$ as the set of all behavior patterns of length $l$,  $\mathcal{S}_l = \{ b_1 \circ b_2 \circ ... \circ b_l | b_i \in \mathcal{B} \} $.  
And we define $\textbf{N}_{u, i}^R$ as the number of paths between nodes $u$ and $i$ that have an $R$-type behavior pattern.

\subsection{Behavior pattern-based recommendation}

In this section, we will elaborate on our target behavior recommendation based on the observed behavior patterns between users and items. 
We harness these behavior patterns as features for our prediction model, with the presence of direct target behavior interactions between users and items $\textbf{E}^{|\mathcal{B}|}$ as labels. 
This transforms the multi-behavior recommendation task into a pattern-based prediction task.
We define $\mathcal{T}$ to represent all types of behavior patterns considered by the algorithm. 
It's important to note that direct interactions corresponding to the target behavior are intentionally omitted from $\mathcal{T}$ because they are the very outcomes we aim to predict.
Therefore $\mathcal{T}=\bigcup_{x=0}^{\alpha} \mathcal{S}_{2x+1} - \{ b_{|\mathcal{B}|} \}$, and $\alpha$ is a hyperparameter representing the desired path length we are concerned with, $b_{|\mathcal{B}|}$ is the target behavior.
The algorithm takes the number of paths $\textbf{N}_{u, i}$, which corresponds to the behavior patterns within $\mathcal{T}$ between user $u$ and item $i$ as input, and the output is a binary label indicating the presence (1) or absence (0) of the target behavior.
Below we illustrate this transformation with an example.

\begin{table}[t]
\setlength{\abovecaptionskip}{5pt}
\setlength{\belowcaptionskip}{-5pt}
  \centering
  \tiny
  \caption{Data for the behavior pattern based prediction}
  \label{table:simu-set}
  \resizebox{\linewidth}{!}{
    \begin{tabular}{C{0.6cm}C{0.6cm}C{0.6cm}C{0.6cm}C{0.6cm}C{0.6cm}C{0.6cm}C{0.6cm}}
    \toprule

\multirow{2}*{User} & \multirow{2}*{Item}  & \multicolumn{5}{c}{Features}   & \multicolumn{1}{c}{Label}   \\
\cmidrule(lr){3-7}  

&  &  $v$ & $c$ & $v \circ v \circ v$ & $v \circ v \circ c$ & $\dots$ & $p$  \\ \midrule

$u_3$ & $i_1$ & 0 & 0 & 1 & 2 & $\dots$ & 0 \\
$u_3$ & $i_2$ & 1 & 0 & 0 & 0 & $\dots$ & 0 \\
$u_3$ & $i_3$ & 1 & 0 & 0 & 0 & $\dots$ & 1 \\

    \bottomrule
    \end{tabular}
  }
\end{table}

Consider an interaction graph as shown in Figure 1, we abbreviate Purchase as $p$, Cart as $c$, and Page View as $v$,
then the data for the behavior pattern prediction task between user $u_3$ and items $i_1$, $i_2$, and $i_3$ is shown in Table 1.
Due to the direct Page View interaction between $u_3$ and $i_3$, therefore $\textbf{N}_{u_3,i_3}^v = 1$.
Moreover, as there are two paths (
$u_3 \stackrel{v} \rightarrow i_2  \stackrel{v} \rightarrow u_2 \stackrel{c} \rightarrow i_1$
and 
$u_3 \stackrel{v} \rightarrow i_2  \stackrel{v} \rightarrow u_1 \stackrel{c} \rightarrow i_1$
) of the $v \circ v \circ c$ pattern between $u_3$ and $i_1$, therefore $\textbf{N}_{u_3,i_1}^{v\circ v \circ c} = 2$. Lastly, due to the direct Purchase interaction between $u_3$ and $i_3$, therefore the label between $u_3$ and $i_3$ is 1.

In BPMR, 
we use the Bayesian method to calculate the proportion $\textbf{O}$ between label 1 and label 0, where
\begin{footnotesize}
\begin{align}
    &  \textbf{O}_{u,i} = \frac{P(Y=1|\mathcal{T}, \textbf{N}_{u,i})}{P(Y=0|\mathcal{T}, \textbf{N}_{u,i})}
\propto
\frac{\prod_{f \in \mathcal{T}} P(f|Y=1)^{\textbf{N}_{u,i}^f} }
{\prod_{f \in \mathcal{T}} P(f|Y=0)^{\textbf{N}_{u,i}^f}} ,
\\
&P(f|Y=1) = \frac{\sum_{u \in \mathcal{U}} \sum_{i \in \mathcal{I}} \textbf{N}_{u,i}^f \textbf{E}^{|\mathcal{B}|}_{u,i}}
{ \sum_{u \in \mathcal{U}} \sum_{i \in \mathcal{I}} \sum_{s \in \mathcal{T}} \textbf{N}_{u,i}^s \textbf{E}^{|\mathcal{B}|}_{u,i} } , \\
&P(f|Y=0) = \frac{\sum_{u \in \mathcal{U}} \sum_{i \in \mathcal{I}} \textbf{N}_{u,i}^f (1 - \textbf{E}^{|\mathcal{B}|}_{u,i})}
{ \sum_{u \in \mathcal{U}} \sum_{i \in \mathcal{I}} \sum_{s \in \mathcal{T}} \textbf{N}_{u,i}^s (1 - \textbf{E}^{|\mathcal{B}|}_{u,i}) } ,
\end{align}
    
\end{footnotesize}
where $P(f|Y=1)$ and $P(f|Y=0)$ respectively represent the probabilities of the occurrence of the path corresponding to behavior pattern $f$ when the label is $1$ or $0$.
Finally, based on the proportion $\textbf{O}$, we sort the user's candidate items to generate a recommendation list.

A major challenge in behavior pattern-based recommendation is the high time cost associated with mining behavior patterns. 
However, this process can be transformed into matrix operations, which are acceleratable using GPUs. 
For instance, calculating the number of $b_1 \circ b_2 \circ b_3$ pattern paths between nodes can be efficiently achieved through a simple matrix multiplication, $\textbf{E}^{b_1} {\textbf{E}^{b_2}}^T \textbf{E}^{b_3}$, where $b_i \in \mathcal{B}$.

\begin{table}[t]
\setlength{\abovecaptionskip}{5pt}
\setlength{\belowcaptionskip}{-5pt}
  \centering
  \tiny
  \caption{Properties of Datasets}
  \label{table:simu-set}
  \resizebox{\linewidth}{!}{
    \begin{tabular}{C{1.3cm}C{0.4cm}C{0.4cm}C{1.0cm}C{3.2cm}}
    \toprule
    Datasets  & $\#$User &  $\#$Item & $\#$Interactions & Interaction Behavior Type  \\ \midrule
    Beibei    & 21,716  &  7,977  & 3,338,068  & \{ Page View, Cart, Purchase \}       \\ 
    Taobao    & 48,749  & 39,493  & 1,952,931  & \{ Page View, Cart, Purchase \}     \\ 
    IJCAI-Contest & 17,237  & 33,799  & 811,582  & \{ Page View, Favorite, Cart, Purchase \}       \\  
    \bottomrule
    \end{tabular}
  }
\end{table}

\begin{table*}
\setlength{\abovecaptionskip}{5pt}
\setlength{\belowcaptionskip}{-5pt}
\caption{The performance comparison on three datasets.}
\tiny
\centering
\resizebox{\linewidth}{!}{
\begin{tabular}{L{1.0cm}C{0.8cm}C{0.8cm}C{0.8cm}C{0.8cm}C{0.8cm}C{0.8cm}C{0.8cm}C{0.8cm}C{0.8cm}C{0.8cm}C{0.8cm}C{0.8cm}}
\toprule
\multirow{2}*{Algorithm}  & \multicolumn{4}{c}{Beibei}    & \multicolumn{4}{c}{Taobao}   & \multicolumn{4}{c}{IJCAI-Contest}     \\
\cmidrule(lr){2-5}   \cmidrule(lr){6-9}  \cmidrule(lr){10-13}  

& R@10  & R@50 &  N@10  & N@50  & R@10  & R@50 &  N@10  & N@50 & R@10  & R@50 &  N@10  & N@50  \\
\midrule
NGCF      & 0.0294 & 0.1073 & 0.0140 & 0.0298    & 0.0143 & 0.0385 & 0.0074 & 0.0116   & 0.0155 & 0.0263 & 0.0071 & 0.0122   \\
LightGCN  & 0.0555 & 0.1637 & 0.0285 & 0.0520    & 0.0410 & 0.0985 & 0.0226 & 0.0338   & 0.0276 & 0.0672 & 0.0149 & 0.0244   \\
SGL       & 0.0590 & 0.1699 & 0.0305 & 0.0545    & 0.0503 & 0.1223 & 0.0279 & 0.0420   & 0.0262 & 0.0652 & 0.0161 & 0.0237   \\
SimGCL    & 0.0615 & 0.1693 & 0.0303 & 0.0534    & 0.0502 & 0.1235 & 0.0282 & 0.0432   & 0.0285 & 0.0731 & 0.0158 & 0.0267   \\

\midrule

MATN      & 0.0464 & 0.1221 & 0.0266 & 0.0423    & 0.0258 & 0.0749 & 0.0133 & 0.0228   & 0.0126 & 0.0346 & 0.0056 & 0.0087   \\
MBGMN     & 0.0416 & 0.1440 & 0.0233 & 0.0436    & 0.0497 & 0.1312 & 0.0250 & 0.0431   & 0.0307 & 0.0655 & 0.0174 & 0.0262   \\
CML       & 0.0577 & 0.1963 & 0.0285 & 0.0578    & 0.0306 & 0.0921 & 0.0143 & 0.0277   & 0.0232 & 0.0491 & 0.0143 & 0.0185   \\
EHCF      & 0.1532 & 0.3326 & 0.0808 & 0.1206    & 0.0712 & 0.1611 & 0.0401 & 0.0594   & 0.0267 & 0.0543 & 0.0148 & 0.0219   \\ 
S-MBRec   & 0.1705 & 0.3697 & 0.0865 & 0.1316    & 0.0814 & 0.1887 & 0.0448 & 0.0674   & 0.0285 & 0.0812 & 0.0147 & 0.0276   \\
GHCF      & 0.1933 & 0.3784 & 0.1011 & 0.1431    & 0.0817 & 0.1902 & 0.0435 & 0.0672   & 0.0315 & 0.0839 & 0.0172 & 0.0288   \\

MBSSL     
& \underline{0.2235} & \underline{0.3811} & \underline{0.1271} & \underline{0.1626}    
& \underline{0.1027} & \underline{0.2108} & \underline{0.0583} & \underline{0.0818}   
& \underline{0.0399} & \underline{0.0964} & \underline{0.0207} & \underline{0.0329}   \\

\midrule
BPMR                  
& \textbf{0.4887} & \textbf{0.6529} & \textbf{0.2500} & \textbf{0.2898} 
& \textbf{0.4025} & \textbf{0.6481} & \textbf{0.2233} & \textbf{0.2803} 
& \textbf{0.1971} & \textbf{0.5696} & \textbf{0.0961} & \textbf{0.1794} \\
\midrule

\textit{Improv.}       
& 118.66\% & 71.32\% & 96.70\% & 78.23\% 
& 291.92\% & 207.45\% & 283.02\% & 242.67\%
& 393.98\% & 490.87\% & 364.25\% & 445.29\%
\\
\bottomrule
\end{tabular}
}
\end{table*}

\section{Experiments}



\subsection{Experimental Setting}


\subsubsection{Datasets.}
We conduct experiments on three public datasets \textit{i.e.},
Beibei,
Taobao,
IJCAI-Contest.
Table 2 lists the details of these three datasets.
1) \textbf{Beibei.}
The dataset originates from Beibei, a major online retailer of infant products in China  \cite{xiaGraphMetaNetwork2021,chenGraphHeterogeneousMultiRelational2021,xuMultibehaviorSelfsupervisedLearning2023}. 
2) \textbf{Taobao.}
This dataset from Taobao, a leading e-commerce platform in China \cite{xiaGraphMetaNetwork2021,xuMultibehaviorSelfsupervisedLearning2023}.
3) \textbf{IJCAI-Contest.}
The dataset, released by IJCAI competition and sourced from a B2C retail system \cite{xiaGraphMetaNetwork2021,xuMultibehaviorSelfsupervisedLearning2023}.


\subsubsection{Baselines.}
In addition to the proposed BPMR, we implemented the following state-of-the-art single-behavior and multi-behavior models as baselines.
The single-behavior models include:
\textbf{NGCF} \cite{wangNeuralGraphCollaborative2019},
\textbf{LightGCN} \cite{heLightGCNSimplifyingPowering2020},
\textbf{SGL} \cite{wuSelfsupervisedGraphLearning2021},
\textbf{SimSGL} \cite{yuAreGraphAugmentations2022}.
The multi-behavior models include:
\textbf{MATN} \cite{xiaMultiplexBehavioralRelation2020},
\textbf{MBGMN} \cite{xiaGraphMetaNetwork2021},
\textbf{CML} \cite{weiContrastiveMetaLearning2022},
\textbf{EHCF} \cite{chenEfficientHeterogeneousCollaborative2020},
\textbf{S-MBRec} \cite{guSelfsupervisedGraphNeural2022},
\textbf{GHCF} \cite{chenGraphHeterogeneousMultiRelational2021},
\textbf{MBSSL} \cite{xuMultibehaviorSelfsupervisedLearning2023}.

\subsubsection{Evaluation Metrics.}
Based on previous research, we adopt the widely used leave-one-out evaluation method and utilize two ranking metrics: \textit{Recall@K} and \textit{NDCG@K}.
Like MBSSL \cite{xuMultibehaviorSelfsupervisedLearning2023}, we rank all items for each user, excluding positive ones.

\subsubsection{Parameter Settings.}
Considering both performance and cost, 
we have set the value of hyperparameter $\alpha$ to 1, 
focusing on behavior patterns of lengths 1 and 3. 
Based on statistics from the Taobao dataset, an average user's 3-hop neighbors can cover $97.12\%$ of items, ensuring the richness of recommendations. 
For the features extracted from each behavior pattern, we normalize them using the z-score normalization method.

\begin{figure}[t]
\setlength{\abovecaptionskip}{0pt}
\setlength{\belowcaptionskip}{-5pt}
\centering
\includegraphics[width=0.4\textwidth]{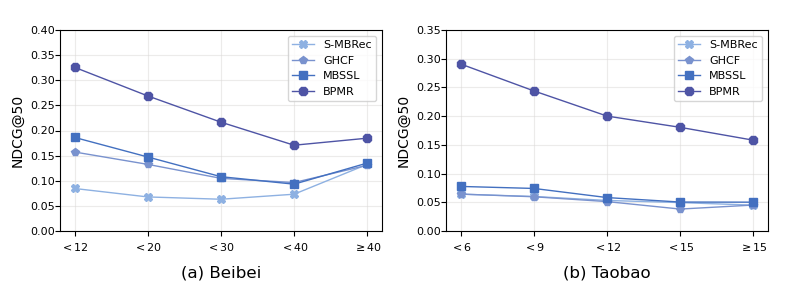}
\caption{Performance on sparse target behavior. }
\label{fig:label}
\end{figure}

\begin{figure}[t]
\setlength{\abovecaptionskip}{0pt}
\setlength{\belowcaptionskip}{-5pt}
\centering
\includegraphics[width=0.4\textwidth]{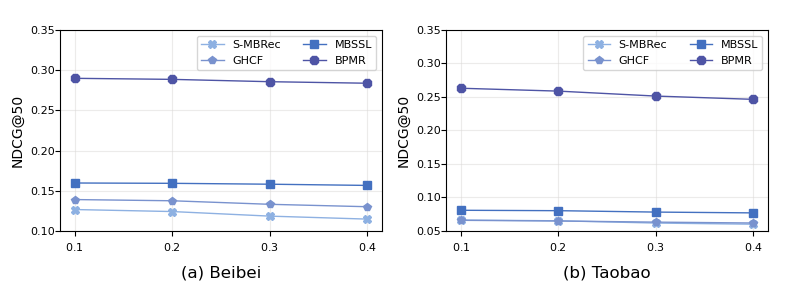}
\caption{Performance on noisy auxiliary behaviors.}
\label{fig:label}
\end{figure}

\subsection{Experimental Results and Analysis}
\subsubsection{Overall Performance.}

In this section, we compare our BPMR's performance with baseline models across three datasets. 
Table 3 summarizes the results. For top K = $\{10, 50\}$ in Recall and NDCG, BPMR significantly outperforms the second-best model, showing average improvements of $94.99\%$ and $87.47\%$ on Beibei, $249.69\%$ and $262.85\%$ on Taobao, and $442.43\%$ and $404.77\%$ on IJCAI-Contest datasets. 
Additionally, BPMR's performance on the Taobao dataset is notably better than on Beibei, and it shows further significant improvement on the IJCAI-Contest dataset compared to Taobao.
BPMR excels due to two key factors: (1) it directly examines node behavior patterns to overcome the limitations of graph neural networks in representing specific behaviors and accurately capturing user preferences, and (2) it leverages the extensive item variety of the Taobao dataset and the broad interaction categories of the IJCAI-Contest dataset, thereby revealing richer behavior patterns and enhancing performance on both datasets.

\subsubsection{Performance on sparse target behavior.}

Based on the degree of the user on the purchase behavior, we divided the test users into five groups and calculated the average performance of each group under the NDCG@50 metric. As shown in Figure 3, the experimental results demonstrate that the BPMR consistently outperforms other models across users with different levels of sparsity. 
This is due to BPMR using path proportions reflecting different behavior patterns. 
Despite changes in node degrees, these path proportions and behavior patterns remain consistent. 

\subsubsection{Performance on noisy auxiliary behaviors.}

We introduced varying levels of random noise ($10\%$, $20\%$, $30\%$, $40\%$) into the auxiliary behavior data to simulate real-world uncertainties and their impact on recommender system performance. 
Results, shown in Figure 4, reveal that the BPMR consistently outperforms the current best baseline models across datasets with different noise proportions. 
This is because, although noise affects behavior patterns differently, BPMR adjusts the weights of these behavior patterns during recommendation, maintaining robust performance 
across datasets with varying noise levels.

\section{Conclusions}

In this paper, we provide a Behavior Pattern Mining-based Multi-behavior Recommendation (BPMR) algorithm that analyzes behavior patterns between nodes, using a Bayesian approach for recommendations. This method overcomes GNNs' shortcomings in depicting specific behavior patterns and user preferences. 
Extensive experiments validate the effectiveness of our model.

\section{ACKNOWLEDGMENTS}
This work is sponsored by 
the National Natural Science Foundation of China (Nos. 61973180, 62172249, 62202253, 62272254), 
the Natural Science Foundation of Shandong Province (Nos. ZR2021MF092, ZR2021QF074, ZR2019MF014, ZR2019MF033, ZR2022MF326)
and
the China Scholarship Council (Grant No. 202308370301).


\bibliographystyle{ACM-Reference-Format}
\bibliography{BPMRref.bib}

\end{document}